\documentclass[doublecol]{epl2} 

\begin{document}

\title{Bound states of moving potential wells in discrete wave mechanics}
\shorttitle{Rapidly-oscillating scatteringless  ... } 

\author{S. Longhi \inst{1,2}}
\shortauthor{S. Longhi}

\institute{                    
  \inst{1}  Dipartimento di Fisica, Politecnico di Milano, Piazza L. da Vinci 32, I-20133 Milano, Italy\\
  \inst{2}  Istituto di Fotonica e Nanotecnlogie del Consiglio Nazionale delle Ricerche, sezione di Milano, Piazza L. da Vinci 32, I-20133 Milano, Italy}
\pacs{03.65.-w}{Quantum mechanics}
\pacs{03.65.Ge} {Solutions of wave equations: bound states}
\pacs{03.65.Nk}{Scattering theory}

\abstract{Discrete wave mechanics describes the evolution of classical or matter waves on a lattice, which is governed by a discretized version of the Schr\"odinger  equation. While for a vanishing lattice spacing wave evolution of the continuous Schr\"odinger  equation is retrieved, spatial discretization and lattice effects can  deeply modify wave dynamics. Here we discuss implications of breakdown of exact Galilean invariance of the discrete Schr\"odinger  equation on the bound states sustained by a smooth potential well which is uniformly moving on the lattice with a drift velocity $v$. While in the continuous limit the number of bound states does not depend on the drift velocity $v$, as one expects from the covariance of ordinary Schr\"odinger  equation for a Galilean boost, lattice effects can lead to  a larger number of bound states for the moving potential well as compared to the potential well at rest. Moreover, for a moving potential bound states on a lattice become rather generally quasi-bound (resonance) states.} 
\maketitle

\section{Introduction}
One of the cornerstones of non-relativistic quantum mechanics is the Schr\"odinger  equation, which describes the temporal evolution of a particle wave function based on its initial state. Traditionally, in wave mechanics space and time are considered continuous. However, on several occasions authors have debated about the nature of space-time manifold and the possibility of considering variant forms of the  Schr\"odinger  equation in which the wave function is defined on discrete lattice sites of space, time, or space-time, instead of on the space-time continuum \cite{r1,r2,r3,r4,r5,r6,r7,r8,r9,r10,r11,r12}. Fundamental limits to a minimum measurable length were suggested in the early days of quantum physics, notably by Heisenberg, and appear, for example, in modern theories of loop quantum gravity, where space-time looks granular \cite{r13}. A phenomenological approach to account for a minimum length scale is to consider an extension of the uncertainty principle by deforming the canonical commutation relations of position and momentum operators \cite{r14}, leading to an extended form of the Sch\"odinger equation \cite{r15,r16}. Other simple models consider the non-relativistic Schr\"odinger equation defined on a discrete lattice \cite{r7,r8,r9,r10,r11,r12,r17,r18,r19,r20,r21,r22}, leading to so-called discrete wave mechanics \cite{r7} or discrete quantum mechanics \cite{r20}.  While earlier models of discrete wave mechanics \cite{r7,r8,r9,r10,r11,r12} did not find great relevance as foundational theories, in several physical contexts, ranging  from condensed-matter physics to chemistry and photonics, the discrete version of the Schr\"odinger equation has demonstrated to provide an excellent description of quantum or classical transport on a lattice \cite{r23,r24,r25,r25bis,r26,r27,r28}. In the limit as the lattice spacing approaches zero, the solution to the discrete Schr\"odinger equation reduces to the one of the continuum theory, however space discretization can give rise to important and noticeable deviations from the continuous equations.  Important effects include negative effective mass,  Bragg reflection and Bloch oscillations \cite{r24,r25,r26,r29}. The discrete version of the Schr\"odinger equation with diagonal (on-site) disorder is also the prototypal model of Anderson localization \cite{r29bis}, whereas the discrete Schr\"odinger equation with a sinusoidal potential describes a crystal electron in a uniform magnetic (Harper model) which yields a fractal structure of energy spectrum (Hofstadter butterfly \cite{r29tris}).\\ 
Another important effect of space discretization, which has not received so far much attention, is breakdown of the covariance of the Schr\"odinger  equation for Galilean boosts \cite{r30}. Such a result has a deep physical consequence and indicates that discrete wave dynamics is distinct in different inertial reference frames. For example, recent works have predicted that reflectionless potentials of the continuous Schr\"odinger equation, such as Kramers-Kronig potentials \cite{r31,r32}, become reflective on a lattice when they are at rest \cite{r33}, while potentials that are reflective at rest can become fully transparent when they move on the lattice at sufficiently high speeds \cite{r34}. This is of course in contrast with the continuous limit of the Schr\"odinger  equation because wave scattering from a potential barrier or well is invariant for a Galilean boost: a potential barrier can not become reflectionless by just observing the scattering process in a moving reference frame!\\
In this Letter we disclose a somewhat paradoxical result of discrete wave mechanics arising from breakdown of the Galilean invariance on a lattice. We consider a smooth potential well on a lattice at rest that sustains some bound states. As the potential well is drifting at some speed $v$ on the lattice, the number of bound states can increase under certain conditions. Such a result \revision{constrasts} with the continuous limit prediction, which obviously requires invariance of wave dynamics (and hence of number of bound states of the Hamiltonian) for observers in relative uniform motion each other. Moreover, the energy of bound states of a moving potential well on the lattice are embedded into the continuous spectrum and thus they are strictly speaking resonance states. This point is illustrated in details by considering the quantum harmonic oscillator on a lattice.

\section{Galilean invariance for continuous and discrete Schr\"odinger equations} 
We consider the time-dependent one-dimensional Schr\"odinger equation on a discrete tight-binding lattice with with a drifting potential $V(x,t)=V(x+vt)$, which can be written in the form \cite{r23,r24,r25,r33}
\begin{equation}
i \hbar \frac{\partial \psi}{\partial t}= -\frac{\hbar^2}{ma^2} [\cos (a \hat{p}_x / \hbar)-1] \psi+ V(x+vt) \psi
\end{equation}
where  $\psi(x,t)$ is the complex wave function amplitude, $a$ is the lattice period, $\hat{p_x}=-i \hbar \partial_x$ is the momentum operator, $m$ is an effective mass, and $v$ is the drift velocity of the potential. An equivalent form of Eq.(1) is given by
\begin{eqnarray}
i \hbar \frac{\partial \psi}{\partial t}  & = &  -\frac{\hbar^2}{2 ma^2} \left[ \psi(x+a,t)+\psi(x-a,t)-2 \psi(x,t)\right]  \nonumber \\
& +  & V(x+vt) \psi(x,t).
\end{eqnarray}
Clearly, the usual continuous Schr\"odinger equation for a quantum particle of mass $m$ is formally obtained from Eq.(1) in the limit $a \rightarrow 0$ by assuming the parabolic approximation $\cos(a \hat{p}_x / \hbar) -1 \simeq -a^2 \hat{p}_x^2/2 \hbar^2$ for the kinetic energy operator. Here we are mainly concerned to investigate the {\em quasi-continuum} (long-wavelength) limit of the discrete Schr\"odinger equation assuming a potential $V(x)$ that varies on a characteristic spatial scale $l$ much longer than the lattice period $a$. Discreteness of the lattice is taken into account by considering higher-order terms in the power series expansion of the kinetic energy operator, i.e. beyond the parabolic approximation. In the continuous limit, it is known that the Schr\"odinger equation is invariant under Galilean boosts \cite{r30}, however in the quasi-continuum limit the invariance is not exact, indicating that the wave dynamics may change depending on the drift velocity $v$ of the potential. To show breakdown of the Galilean invariance for the discrete Schr\"odinger equation in the quasi-continuum limit, let us consider the Galilean transformation of coordinate system
\begin{equation}
x^{\prime}=x+vt \;, \;\; t^{\prime}=t
\end{equation}
so that Eq.(2) takes the form
\begin{eqnarray}
i \hbar \frac{\partial \psi}{\partial t^{\prime}}  & = &  -\frac{\hbar^2}{2 ma^2} \left[ \psi(x^{\prime}+a,t^{\prime})+\psi(x^{\prime}-a,t^{\prime})-2 \psi(x^{\prime},t^{\prime})\right]  \nonumber \\
& +  & V(x^{\prime}) \psi(x^{\prime},t^{\prime})-i v \hbar \frac{\partial \psi}{\partial x^{\prime}}.
\end{eqnarray}
After introduction of the gauge transformation for the wave function
\begin{equation}
\psi(x^{\prime},t^{\prime})= \psi^{\prime}(x^{\prime},t^{\prime}) \exp \left(- i q x^{\prime}-i \gamma t^{\prime} \right)
\end{equation}
where $q$ and $\gamma$ are defined by the following relations
\begin{eqnarray}
\sin (q a) = \frac{mva}{\hbar} \\
\gamma = -vq-\frac{\hbar}{ma^2} \left[ \cos (qa)-1\right]
\end{eqnarray}
from Eq.(4) one readily obtains the following modified Schr\"odinger equation in the $(x^{\prime},t^{\prime})$ reference frame
\begin{equation}
i \hbar \frac{\partial \psi^{\prime}}{\partial t^{\prime}}=-\frac{\hbar^2}{2 m^*} \left( \frac{\partial^2}{\partial x^{\prime 2}}+ \frac{m^*}{m} \mathcal{P} \right) \psi^{\prime}+ V(x^{\prime}) \psi^{\prime}
\end{equation}
where $m^{*}$ is a renormalized mass, given by 
\begin{equation}
m^{*} \equiv \frac{m}{\cos(qa)}= \frac{m}{\sqrt{1-(mva / \hbar)^2}}
 \end{equation}
 and the operator $\mathcal{P}$ is defined by
 \begin{eqnarray}
 \mathcal{P} & = & a \left[ 2 \cos (qa) \sum_{n=2}^{\infty} \frac{a^{2n-3}}{(2n) !}  \frac{\partial^{2n}}{\partial x^{\prime 2n}}  \right. \nonumber \\
 & - & \left. 2 i \sin (qa) \sum_{n=1}^{\infty} \frac{a^{2n-2}}{(2n+1) !}  \frac{\partial^{2n+1}}{\partial x^{\prime 2n+1}} \right] .
 \end{eqnarray}
 Clearly Eq.(8), which is an {\em exact} equation, shows that  the discrete Schr\"odinger equation is not covariant for a Galilean boost, even in the long-wavelength (quasi-continuum) limit. 
 The continuous limit of ordinary Schr\"odinger equation is retrieved by letting $a \rightarrow 0$ in the above equations. In this limit, from Eqs.(6), (7) and (9) one has $q \simeq mv / \hbar$, $\gamma \simeq -mv^2 / ( 2 \hbar)$, $m^* \simeq m$, and the correction term $\mathcal{P}$ to the kinetic energy in Eq.(8), which scales like $\sim a$ [Eq.(10)], can be neglected. Therefore one obtains 
 \begin{equation}
 i \hbar \frac{\partial \psi^{\prime}}{\partial t^{\prime}}=-\frac{\hbar^2}{2m} \frac{\partial^2 \psi^{\prime}}{\partial x^{\prime 2}}+V(x^{\prime}) \psi^{\prime}
 \end{equation}   
 indicating exact invariance of the Schr\"odinger equation for a Galilean boost \cite{r30}. However we can consider another limiting case, corresponding to the double limit of a small lattice period $a \rightarrow 0$ and a fast speed of the moving potential $v \rightarrow \infty$, with 
 \begin{equation}
 \frac{mva}{ \hbar}
 \end{equation}
  finite and smaller than one. In this double limit, the correction term $\mathcal{P}$ to the kinetic energy in Eq.(8), which scales like $\sim a$ regardless of the strength of the drift velocity $v$, can be neglected, however the renormalized mass $m^*$ differs from $m$ and Eq.(8) takes the form
 \begin{equation}
  i \hbar \frac{\partial \psi^{\prime}}{\partial t^{\prime}}=-\frac{\hbar^2}{2m^*} \frac{\partial^2 \psi^{\prime}}{\partial x^{\prime 2}}+V(x^{\prime}) \psi^{\prime}
 \end{equation} 
  In this limit the invariance of the wave equation for a Galileian boost is not exact since in different inertial reference frames the wave equation is a continuous Schr\"odinger equation but with a renormalized mass which depends on the drift velocity $v$ [Eq.(9)]. Note that the renormalized mass $m^*$ in the moving reference frame is larger 
  than the mass $m$ in the rest reference frame.
 
\section{Bound states of a moving potential well} Let us consider a potential well $V(x)$, with $V(x) \rightarrow 0$ as $|x| \rightarrow \infty$, sustaining $N$ bound states with negative energies in the continuous limit. Owing to Galilean invariance, in this limit a drift of the potential well does not change the number of bound states. However, space discretization breaks Galileian invariance, and at leading order discreteness can be accounted for by the introduction of a velocity-dependent renormalization of mass as discussed in the previous section [Eqs.(9) and (13)]. Since $m^* >m$, even though mass renormalization is a small effect it might happen that the number of bound states for the moving potential well becomes {\em larger} than the one of the same potential well at rest.  This is a somewhat paradoxical result that arises when one try to extend ordinary non-relativistic quantum mechanics by the introduction of  a minimum length (space discretization) into the Schr\"odinger equation \footnote{In earlier papers dealing with discrete quantum mechanics (see e.g. \cite{r7,r8}), the spatial length $a$ of the lattice is assumed of the order of the Compton wavelength, i.e. $a=\hbar /(mc)$, which follows from the Heisenberg uncertainty principle $ \Delta x \Delta p \sim \hbar$ by taking $\Delta x=a$ and an upper limit for the uncertainty of momentum $\Delta p \sim mc$ requested by special relativity. Interestingly, for $a=\hbar /(mc)$ the renormalized mass $m^*$, given by Eq.(9), takes the form of the relativistic mass in special relativity, namely $m^*=m/\sqrt{1-(v/c)^2}$. Note that for a drift velocity $v$ much smaller than $c$, i.e. in the non-relativistic limit, the correction $m^*$ to $m$ is a small effect.}. To clarify this point, let us consider the stationary Schr\"odinger equation with renormalized effective mass in the moving reference frame. By setting $\psi^{\prime}(x^{\prime},t^{\prime})=\varphi^{\prime} (x^{\prime}) \exp(-i E t^{\prime} / \hbar)$ in Eq.(13), one obtains
\begin{equation}
E^{\prime} \varphi^{\prime}(x^{\prime})=-\frac{\hbar^2}{2 m} \frac{d^2 \varphi^{\prime}}{ d x^{\prime 2}}+V^{\prime}(x^{\prime}) \varphi^{\prime}(x^{\prime})
\end{equation}
where $E^{\prime} \equiv E (m^*/m)$, $E$ is the energy eigenvalue, and
\begin{equation}
V^{\prime}(x^{\prime})=\frac{m^*}{m} V(x^{\prime}).
\end{equation}
 Bound states correspond to normalizable eigenfunctions of Eq.(14) with negative energy eigenvalues. For the sake of definiteness, let us assume as an example the P\"oschl-Teller potential well
\begin{equation}
V(x^{\prime})=-\frac{\hbar^2}{2ml^2} \frac{\nu(\nu+1)}{\cosh^2(x^{\prime}/l)}
\end{equation}
($ \nu,l>0$ real), which is exactly solvable \cite{r35}; the results, however, can be extended rather generally to any other shape of the potential well. For the potential well at rest, i.e. $v=0$, $m^*=m$ and $V^{\prime}(x^{\prime})=V(x^{\prime})$, there are $N=1+[ \nu]$ bound states with energies $E=E_n=- \hbar^2 (\nu-n)^2/(2 ml^2)$ ($n=0,1,2,..., [ \nu ]$), where $[ \nu]$ is the integer part of $\nu$ \cite{r35}. For $\nu$ integer, the potential is reflectionless and there is one just unbound state with zero energy. Let us assume that $\nu$ is close to an integer $N$ from below, i.e. $\nu=N^-$, so that the potential well at rest sustains $N$ bound states and just one unbound state. As the potential well drifts with a speed $v$, from Eq.(15) it readily follows that the depth of the potential well is effectively increased by the (generally small) amount $m^*/m$, which is equivalent to replace $\nu \rightarrow \nu^*$ in the P\"oschl-Teller potential (16), with $\nu^*(\nu^*+1)=(m^*/m) \nu( \nu+1)$. Since $\nu^*=N^+$, for the moving potential well the stationary Schr\"odinger equation (14) shows that there are now $(N+1)$ bound states. Such a result is confirmed by direct numerical simulations of the discrete Schr\"odinger equation (2), which are illustrated in the next section.\\
The above analysis indicates that, at leading order and for a vanishingly small lattice period, the main effect of space discretization for a moving potential well can be accounted for by a simple renormalization of mass $m$ in the continuous Schr\"odinger equation [Eq.(13)], however one should ask whether the correction term $\mathcal{P}$ of the kinetic energy operator [Eqs.(8) and (10)] can play some role, even though it is a small (perturbative) term in the equation as $ a \rightarrow 0$. To this aim, let us consider plane wave solutions to Eq.(8) in the absence of the potential well, which should asymptotically reproduce, as $|x^{\prime}| \rightarrow \infty$, the scattering states  
of the potential well in the moving reference frame. It can be readily shown that the plane wave $\psi^{\prime}(x^{\prime},t^{\prime})= \exp [-ik x^{\prime}-i E(k) t^{\prime} / \hbar ]$ with wave number $k$ is a solution to Eq.(8) when $V=0$, with the energy dispersion curve $E=E(k)$ given by
\begin{equation}
E(k)= \frac{2 \hbar^2}{ma^2} \sin \left( \frac{k+2q}{2}a \right) \sin \left( \frac{ka}{2}\right) -\hbar v k.
\end{equation}
Note that, while in the continuous limit $a \rightarrow 0$ $E(k)$ can be approximated by the parabolic curve $E(k) \simeq (\hbar^2 / 2m) k(k+2q)-\hbar vk$, which is limited from below, for a finite (though small) lattice period $E(k)$ is unbounded both from above and below whenever $v \neq 0$. This means that the energies of bound states of the potential well $V(x^{\prime})$ in the moving reference frame $(x^{\prime},t^{\prime})$, determined at leading order as eigenvalues of Eq.(14), are embedded into the continuous energy spectrum of scattered states. Therefore, strictly speaking the bound states of a moving potential well on a lattice become resonance states when the perturbation term $\mathcal{P}$ is included, i.e. there are not any truly bound states for a moving potential well on a lattice. This point can be illustrated by considering in details the case of a parabolic potential well, i.e. the quantum harmonic oscillator on a lattice \cite{r25,r36} $V(x^{\prime})=\frac{1}{2} m \omega^2 x^{\prime 2}$ [Fig.1(a)]. An exact bound state of the drifting harmonic oscillator on the lattice of energy $E$ can be found by looking for a solution to Eq.(4) of the form $\psi(x^{\prime},t^{\prime})=\varphi(x^{\prime}) \exp(-i E t^{\prime} / \hbar)$, with
\begin{eqnarray}
E \varphi(x^{\prime}) & = & -\frac{\hbar^2}{2ma^2} \left[ \varphi(x^{\prime}+a)+\varphi(x^{\prime}-a)-2 \varphi(x^{\prime})   \right] \nonumber \\
& + & \frac{1}{2}m \omega^2 x^{\prime 2} \varphi (x^{\prime})-i v \hbar \frac{d \varphi}{d x^{\prime}}.
\end{eqnarray}
For a bound state, the condition $\int_{-\infty}^{\infty} dx^{\prime} |\varphi(x^{\prime})|^2 < \infty$ should be satisfied. However, one can readily prove that any solution to Eq.(18) is not normalizable, regardless of the value of energy $E$. To this aim, let us introduce the Fourier transform $\hat{\varphi}(k)= \int_{-\infty}^{\infty} d x^{\prime} \varphi(x^{\prime}) \exp(ikx^{\prime})$ of $\varphi(x^{\prime})$, i.e. let us consider the momentum representation of the wave function. In momentum space, Eq.(18) takes the form
\begin{equation}
-\frac{1}{2} m \omega^2 \frac{d \hat{\varphi}^2}{dk^2}+ W(k) \hat{\varphi}=E \hat{\varphi}(k)
\end{equation}
where we have set
\begin{equation}
W(k)=\frac{\hbar^2}{ma^2} \left[ 1- \cos (ka) \right] - \hbar v k. 
\end{equation}
Interestingly, Eq.(19) can be viewed as a continuous stationary Schr\"odinger equation in a potential $W(k)$. 
 If for some energy $E$ $\varphi(x^{\prime})$ were a bound state, then $\int_{-\infty}^{\infty} dk |\hat{\varphi}(k)|^2 < \infty$, i.e.  $\hat{\varphi}(k)$ would be a normalizable (bound) state with energy $E$ for the stationary continuous Schr\"odinger equation (19). Clearly, in the limit $a \rightarrow 0$ the potential $W(k)$, given by Eq.(20), reduces to the shifted parabolic potential $W(k) \simeq \hbar^2k^2/(2m)-\hbar v k$, and the set of energy levels $E=E_n=\hbar \omega (n+1/2)-mv^2/2$ ($n=0,1,2,3,....$)  correspond to Gauss-Hermite bound states of the ordinary quantum harmonic oscillator. However, for a non vanishing (though arbitrarily small) value of $a$ and for a drifting potential ($v \neq 0$), $W(k)$ is unbounded both from above and below [Fig.1(b)], so that the bound states at the energy levels $E_n$ become resonance states, i.e. they are decaying states with some finite (though long) lifetime [see Fig.1(b)] \footnote{As $a \rightarrow 0$ and $W(k)$ approaches the parabolic potential, the lifetime of resonance states diverges and the limit of bound states is retrieved.}. This means that the energy spectrum of Eq.(19) is absolutely continuous and there are not normalizable states.

\begin{figure}
\onefigure[width=7.8cm]{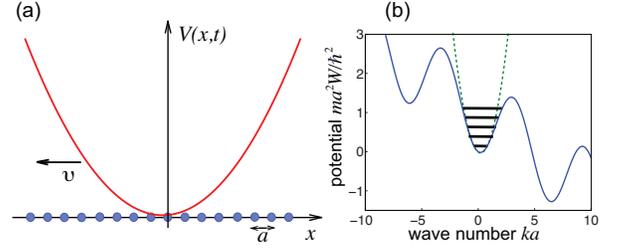}
\caption{(Color online) (a) Schematic of a quantum harmonic oscillator drifting on a lattice. (b) Behavior of the potential $W(k)$ (solid curve) in momentum space representation [Eqs. (19) and (20)] for $mav/ \hbar=0.2$. The dashed line corresponds to the parabolic approximation of the potential. In such a limiting case a set of Gauss-Hermite bound states are sustained by the potential at energies schematically depicted by the thick horizontal segments. Beyond the parabolic approximation, the bound states become resonance states with a finite lifetime.}
\end{figure}

\section{Numerical simulations} 
We checked the main predictions of the theoretical analysis by direct numerical simulations of the discrete 
Schr\"odinger equation (2). To this aim, let us introduce the complex amplitudes $c_n(t)$ of the wave function $\psi$ at the lattice sites $x=na$ by setting
\begin{equation}
c_n(t)= \psi(x=na,t) \exp \left(i \frac{\hbar t}{ma^2} \right)
\end{equation}
with $n= 0 , \pm1, \pm2, ...$. Substitution of Eq.(21) into Eq.(2) yields the following set of coupled equations for the complex amplitudes $c_n(t)$
\begin{equation}
i \frac{dc_n}{dt}= -\kappa (c_{n+1}+c_{n-1})+ \mathcal{V}(na+vt)c_n \equiv \sum_{m} \mathcal{H}_{n,m}c_m
\end{equation}
where we have set $\kappa \equiv \hbar / (2 m a^2)$, $\mathcal{V}(x) \equiv (1 / \hbar) V(x)$ and $\mathcal{H}_{n,m}=-\kappa( \delta_{n,m+1}+\delta_{n,m-1})+\mathcal{V}(na+vt) \delta_{n,m}$.\\
In a first set of numerical simulations,  we assumed a shallow P\"oschl-Teller potential well
\begin{equation}
\mathcal{V}(x)=-\frac{\kappa \nu(\nu+1) (a/l)^2}{\cosh^2 (x/l)}
\end{equation}

\begin{figure}
\onefigure[width=7.8cm]{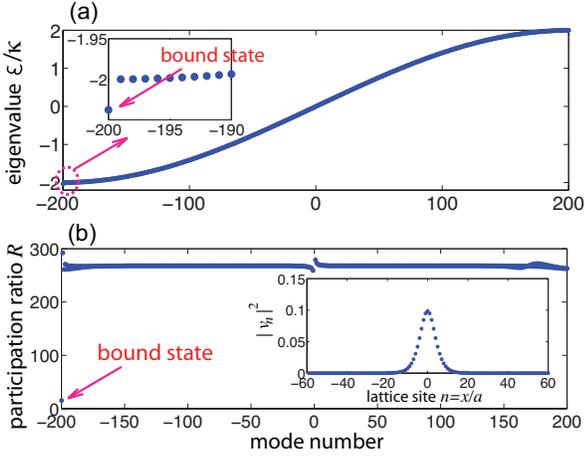}
\caption{(Color online) (a) Numerically-computed behavior of the eigenvalues $\mathcal{E}$ (energy spectrum) of the Hamiltonian matrix $\mathcal{H}_{n,m}$ for the P\"oschl-Teller potential well [Eq.(23)] in a lattice comprising 401 sites with open boundary conditions, and (b) corresponding behavior of participation ratio $R$ of eigenvectors. Parameter values are $a/l=0.2$ and $\nu=0.97$. The potential well sustains one bound state. The inset in (a) shows an enlargement of the energy spectrum near the bottom of the tight-binding sinusoidal band, where the energy of the bound state appears just below the continuous spectrum of extended Bloch waves.  The inset in (b) shows the profile $|v_n|^2$ of the bound state.}
\end{figure}

\begin{figure}
\onefigure[width=7.8cm]{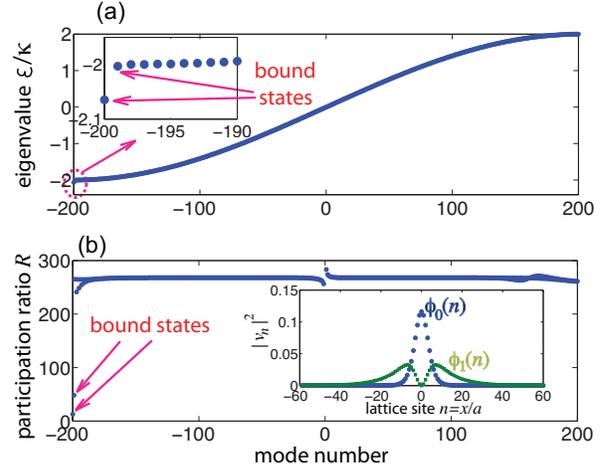}
\caption{(Color online) Same as Fig.2, but for $\nu=1.27$. In this case the potential well sustains two bound states with opposite parity, $\phi_0(n)$ and $\phi_1(n)$, shown in the inset of panel (b).}
\end{figure}

\begin{figure}
\onefigure[width=8.8cm]{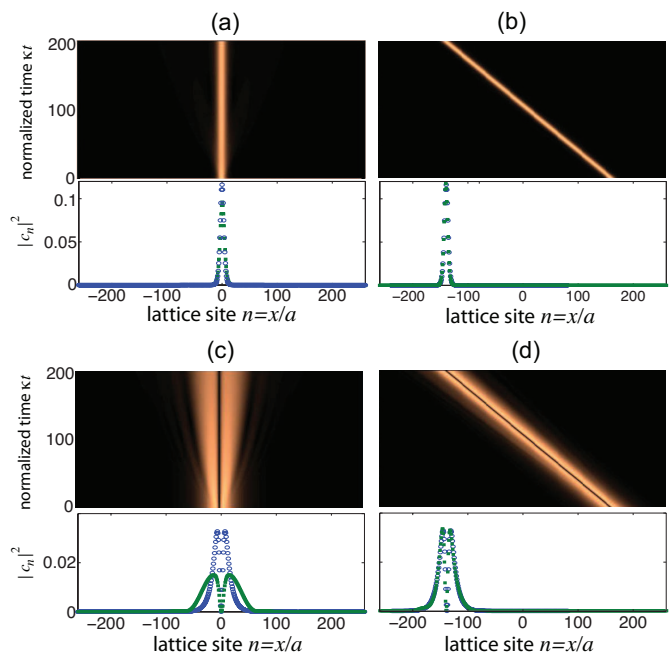}
\caption{(Color online) Numerically-computed evolution of wave packet (snapshots of $|c_n(t)|$ on a pseudo color map) for 
the discrete P\"oschl-Teller potential well (23) and for parameter values $\nu=0.97$ and $a/l=0.2$. In (a) and (c) the potential well is at rest ($v=0$), whereas in (b) and (d) it drifts at the speed  $v=1.5 \kappa a$. The initial wave packet distribution $c_n(0)$, with probability $|c_n(0)|^2$ shown by open circles in the lower insets, is $\phi_0(n)$ in (a), $\phi_1(n)$ in (c),  $\phi_0(n) \exp(iqna)$ in (b) and $\phi_1(n) \exp(iqna)$ in (d), \revision{where $\phi_0(n)$, $\phi_1(n)$ are the even- and odd-parity modes of a well at rest with $\nu=1.27$ shown in the inset of Fig.3(b)} and $qa$ is given by Eq.(6), i.e. $\sin (qa)=v/(2 \kappa a)=0.75$. The detailed behaviors of probability distributions $|c_n(t)|^2$ at final propagation time $\kappa t=200$ are shown by filled squares in the four insets. Note that the P\"oschl-Teller potential well at rest with $\nu=0.97$ cannot localize the initial distribution $\phi_1(n)$ with odd parity, which spreads in time [Fig.4(c)], while localization is possible when the potential well drifts on the lattice [Fig.4(d)].}
\end{figure}
\begin{figure}
\onefigure[width=8.8cm]{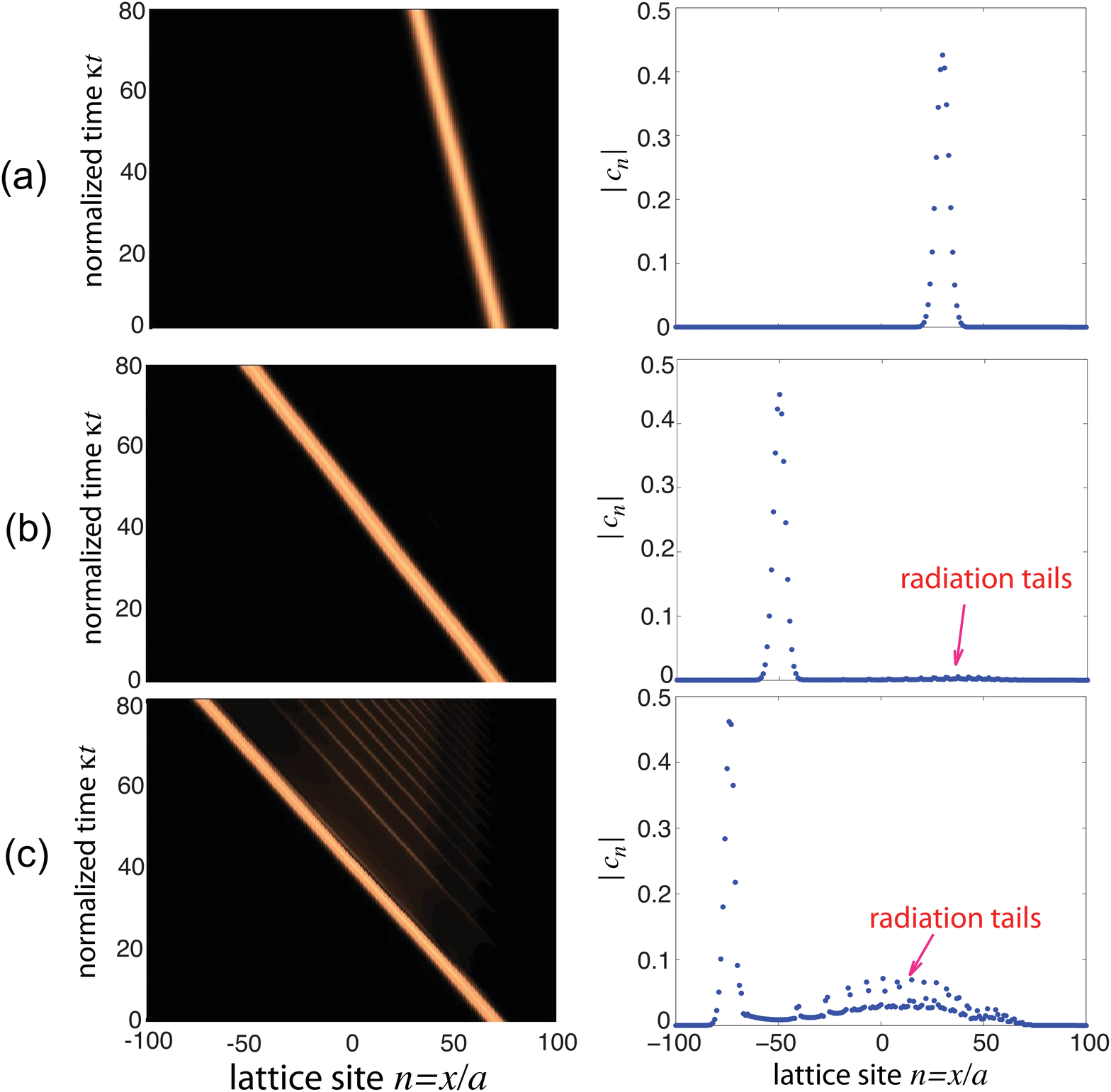}
\caption{(Color online) Numerically-computed evolution of a discrete wave packet in the 
the quantum harmonic oscillator on a lattice for $\Omega a^2=0.02$ and for increasing values of the drift velocity: (a) $ v=0.5 \kappa a$, (b) $v=1.5 \kappa a$, and (c) $v=1.8 \kappa a$. The minimum of the potential well  at initial time $t=0$ is located at the lattice site $n=70$. Left panels show snapshots of $|c_n(t)|$ on a pseudo color map, whereas right panels depict the detailed distribution of the amplitude probability $|c_n|$ at final time $\kappa t=80$. In (b) and (c) radiation tails are clearly observed.}
\end{figure}
\noindent with $a \ll l$. For a potential at rest and in the continuous limit $a / l \rightarrow 0$, the potential well sustains $ N= [ \nu ]+1 $ bound states. We checked such a result by direct numerical computation of the eigenvectors $v_n(\mathcal{E})$ with eigenvalues $\mathcal{E}$ of the matrix $\mathcal{H}_{n,m}$, assuming a finite lattice comprising $(2M+1)=401$ sites ($n=-M,...,M$) with open boundary conditions. The degree of localization of the eigenstate $v_n(\mathcal{E})$ is measured by the participation ratio $R$, given by $R(\mathcal{E})=\left( \sum_n |v_n|^2 \right)^2 / \left( \sum_n |v_n|^4 \right)$ (see, for instance, \cite{r40}). For localized modes, $R \sim 1$  while for extended states $R \sim (2M+1)$. Figures 2 and 3 show, as an example, the numerically-computed behaviors of the energy spectrum and participation ratio versus $\mathcal{E}$ of the discrete P\"oschl-Teller potential well (23) for $a/l=0.2$ and for two values of $\nu$, just below and above $\nu=1$. For $\nu=0.97$ (Fig.2), there is only one bound state with even parity, whereas for $\nu=1.27$ there are two bound states with opposite parity (Fig.3), that we indicate by $\phi_0(n)$ (even-parity or fundamental mode) and $\phi_1(n)$ (odd-parity or excited mode). Note that the fundamental (even parity) mode $\phi_0(n)$ of Fig.3(b) is slightly more confined than the fundamental mode of Fig.2(b)  because of the potential well is slightly deeper in the latter case. To check the increase of bound state number when the potential well drifts on the lattice, we numerically investigated the localization properties of the P\"oschl-Teller potential well with $a/l=0.2$ and $\nu=0.97$, which at rest sustains only one bound state with even parity (Fig.2). The time-dependent coupled equations (22) were integrated using an accurate fourth-order variable-step Runge-Kutta method assuming different initial conditions. The main results are summarized in Fig.4. Wave packet evolution for the P\"oschl-Teller potential well at rest is shown in Figs.4(a) and (c) for two different initial conditions, namely $c_n(0)=\phi_0(n)$ in Fig.4(a) and $c_n(0)=\phi_1(n)$ in Fig.4(c), where $\phi_0(n)$ and $\phi_1(n)$ are the even- and odd-parity states defined above and depicted in the inset of Fig.3(b). Clearly, since the potential well at rest sustains one bound state with even parity, the initial wave packet distribution $\phi_0(n)$, which closely resembles the bound state of the well, remains localized during the evolution, while the initial wave packet distribution $\phi_1(n)$ with odd-parity can not be confined by the potential well and spreads in time, as shown in Fig.4(c). Figures 4(b) and (d) show the wave packet evolution for the potential well drifting on the lattice at the speed $v=1.5 \kappa a$ for the initial conditions $c_n(0)=\phi_0(n) \exp(iqna)$ in (b) and $c_n(0)=\phi_1(n) \exp(iqna)$ in (d), where $qa$ is defined by Eq.(6), i.e. $\sin (qa)=v/(2 \kappa a)=0.75$. Note that in this case both initial wave packet distributions can be confined by the drifting potential well. This result is a clear signature that the drifting potential well effectively sustains two bound states with opposite parity. In fact, for the chosen drift velocity and in the moving reference frame $(x^{\prime}, t^{\prime})$ the potential well appears at rest with an effective depth $\nu^*$ which is increased from $\nu=0.97$ to the value $\nu^*=1.27$. The value $\nu^*$ is obtained from the relation $\nu^*(\nu^*+1)=(m^*/m) \nu( \nu+1)$ derived in the previous section and with the ratio $(m^*/m) \simeq 1.51 $ calculated from Eq.(9). Therefore, the drifting potential well effectively corresponds to the one of Fig.3 and sustains two bound states, which are precisely the two distributions $\phi_{0}(n)$ and $\phi_1(n)$ used as initial conditions.\\
In a second set of simulations, we assumed a parabolic potential well, namely $\mathcal{V}(x)=(1/2) \kappa \Omega x ^2$, which corresponds to the quantum harmonic oscillator on a lattice \cite{r25,r36}. We numerically integrated the time-dependent equations (22) with the initial condition $c_n(0)= \phi_0(n) \exp(iqna)$, where $\phi_0(n)$ is the ground-state (fundamental mode) of the quantum harmonic oscillator, with mass correction term according to Eq.(9), and $q$ is defined by Eq.(6) for a given drift velocity $v$. Figure 5 shows typical behaviors of wave packet evolution for increasing values of the drift velocity $v$. Clearly, as the drift velocity increases one can observe the appearance of radiation tails in the amplitude probability distribution, which are the signature that the bound state decays and is this a resonance state. 

\section{Conclusions} Discrete wave mechanics was proposed as a modified theory of ordinary non-relativistic wave mechanics to phenomenologically describe space discreteness effects \cite{r7,r8,r9,r10,r11,r17,r18,r19,r20}, which are expected to arise at small spatial scales from simple arguments based on special relativity and the Heisenberg uncertainty principle  \cite{r7}. More generally, the discrete version of the Schr\"odinger equation can model transport of classical or quantum waves on a lattice, and space discreteness is essential to explain several major phenomena observed in condensed-matter physics, atom optics and photonics, such as Bloch oscillations, Anderson localization, fractal energy spectra,  etc. While for vanishingly small lattice spacing discrete wave mechanics fully reproduces the ordinary wave dynamics of continuous Schr\"odinger equation, space discreteness manifests itself in a  series of important effects. Here we have disclosed some interesting and somewhat paradoxical results that one would expect in a discrete wave mechanics theory when considering bound states of a moving potential well. Space discreteness breaks exact covariance of the Schr\"odinger  equation for Galilean boosts, making it possible to observe different number of bound states for a potential well depending on its speed. Moreover, the relative motion of the potential well with respect to the lattice unavoidably introduces decay of bound states, which become resonance states. Our results, besides of shading new light onto discrete wave mechanics at a fundamental level, can be of interest in condensed-matter, matter-wave or photonic hopping transport on a lattice. For example, our results show that the number of bound states of a potential on a lattice can be increased by drifting the potential on the lattice, a possibility that would be forbidden in continuous potentials.

\end{document}